\documentstyle[11pt,baaa-eng,twoside,epsf,psfig]{article}
\begin{document}
\vskip 1.0cm
\markboth{C. Scannapieco et al.}{Supernova Feedback in SPH simulations}
\pagestyle{myheadings}


\vspace*{0.5cm}
\parindent 0pt{ COMUNICACI\'ON DE TRABAJO -- CONTRIBUTED  PAPER } 
\vskip 0.3cm
\title{Impact of Supernova Explosions on Galaxy Formation} 

\author{Cecilia Scannapieco}
\affil{Instituto de Astronom\'{\i}a y F\'{\i}sica del Espacio, Buenos Aires, Argentina, cecilia@iafe.uba.ar}

\author{Patricia B. Tissera}
\affil{Instituto de Astronom\'{\i}a y F\'{\i}sica del Espacio, Buenos Aires, Argentina, patricia@iafe.uba.ar}

\author{Simon D.M. White}
\affil{Max-Planck Intstitute for Astrophysics, Garching, Germany, swhite@mpa-garching.mpg.de}

\author{Volker Springel}
\affil{Max-Planck Intstitute for Astrophysics, Garching, Germany, volker@mpa-garching.mpg.de}

\begin{abstract} We study the effects of Supernova (SN) feedback on the formation
of disc galaxies. For that purpose we run simulations using
the extended version of the code {\small GADGET-2}
which includes a treatment of chemical and energy feedback by SN explosions.
We found that our model succeeds in setting a self-regulated
star formation process since an important fraction of the cold gas from the center of the
haloes is efficiently heated up and transported outwards.
The impact of SN feedback on galactic systems is also found to depend on virial mass:
smaller systems are more strongly affected with star formation
histories in which several starbursts can develop. 
Our implementation of SN feedback is also successful in producing
violent outflows of chemical enriched material.
\end{abstract}

\begin{resumen}  
En este trabajo se estudia el impacto de explosiones de Supernova (SN)
en la formaci\'on de galaxias de disco. Para ello se utilizan simulaciones
num\'ericas realizadas con una versi\'on del c\'odigo {\small GADGET-2}
que incluye los efectos de enriquecimiento qu\'{\i}mico e inyecci\'on
de energ\'{\i}a debido a explosiones de SN. Este modelo
es capaz de generar un mecanismo auto-regulado de  formaci\'on
estelar, como resultado del calentamiento eficiente del gas
en el centro de los sistemas y su transporte hacia el halo.
En nuestro modelo, el impacto de las explosiones de SN
en sistemas gal\'acticos depende de la masa virial de los mismos:
los m\'as peque\~nos son afectados m\'as fuertemente
pudiendo presentar varios
brotes de formaci\'on estelar. Nuestro modelo
es tambi\'en capaz de producir
fuertes flujos de gas hacia las regiones externas de las galaxias,
los cuales transportan una importante fracci\'on de metales
hacia  el medio intergal\'actico.

\end{resumen}

\section{Introduction}

Supernova explosions are thought to play a fundamental role
in the evolution of galaxies. Through the ejection of metals
and energy into the interstellar medium, they significantly
affect the gas collapse, the star formation activity
and the chemical patterns. As a consequence, SN feedback would
be able to set a self-regulated mechanism
for the star formation activity and the enrichment of the Universe. Moreover, SN driven outflows
are believed to be responsible for the presence
of metals detected in the intergalactic medium
(e.g. Tytler et al. 1995; Songaila \& Cowie 1996).

In the context of numerical simulations, the treatment of
SN feedback has found severe complications. Several
works have developed models of different complexity
to treat SN feedback in Smoothed Particle Hydrodynamics
(SPH; Gingold \& Monaghan 1977; Lucy 1977) codes (e.g. Katz \& Gunn 1991; 
 Navarro \& White 1993; Metzler \& Evrard 1994;  Marri \& White 2003; Springel \& Hernquist 2003),
finding different levels of success.

\section {Numerical Models}

In this study, we analise simulations of isolated disc galaxies
of different virial masses from idealized initial conditions.
The initial conditions correspond to spherical distributions
of superposed gaseous and dark matter particles, perturbed
to give rise to a density profile of the form $\rho(r)\propto r^{-1}$. The
spheres are initially in solid body rotation with a spin
parameter of $\lambda\approx 0.1$. We have used an initial number
of $9000$ both gaseous and dark matter particles for all masses.

The simulations were performed
with the extended version of the Tree-PM SPH 
code {\small GADGET-2} (Springel \& Hernquist 2002) which
includes a treatment of chemical enrichment and energy feedback
by Supernova (Scannapieco et al. 2005a,b,c). The model is tight to
a multiphase scheme which allows a better description of 
diffuse gas in the context of the SPH technique. 
The interested reader is referred to Scannapieco et al. 2005a,b for
details on the numerical scheme.
All experiments presented  in this study include  chemical enrichment and metal-dependent cooling, and
they all have been run in the context of the multiphase
treatment for the gas component.

It is relevant to this study to mention that the energy feedback
model has only one free input parameter. Within the model,
we define for each star particle with associated SN explosions two
gaseous phases in the following manner: a {\it cold} phase defined as the gas with
$T < 2 T_*$  and $\rho > 0.1 \rho_*$  ($T_* = 4 \times 10^4$ K, 
$\rho_* = 7 \times 10^{-26}$ g cm$^{-3}$) and a {\it hot} phase formed by the remaining gas.
The energy released in the explosions, as well as the metal content,
are distributed into the cold and hot phases:
the cold phase receives a fraction  $\epsilon_c$ while
the hot phase gets the remaining  $\epsilon_h = 1 - \epsilon_c$. Hence, 
the parameter $\epsilon_c$ (or $\epsilon_h$ equivalently) is the
only free parameter to assume.
Note that the value of $\epsilon_c$ is related to the power of the feedback.
However, we note that the effects of varying its value are not completely linear,
owing to the non trivial interplay among energy release, heating and cooling (see
Scannapieco et al. 2005b).

\section {Results}

\begin{figure}  
{\hspace*{2.5cm}   \psfig{figure=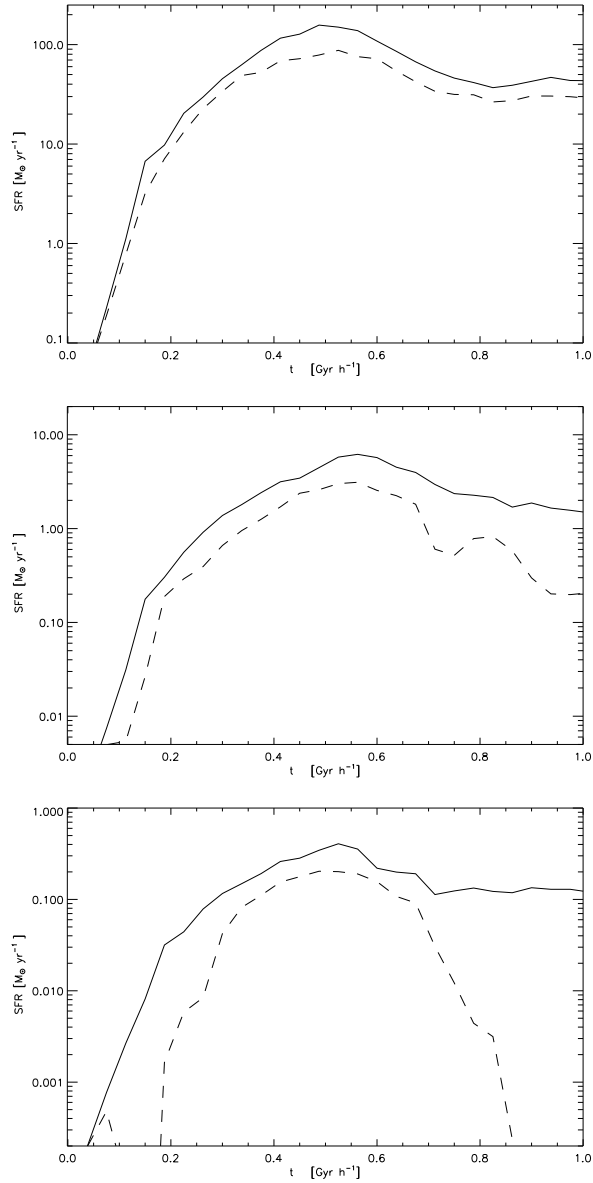,width=8.cm}}
\caption{Evolution of the star formation rates for our tests of
isolated rotating spheres with virial masses of 
$10^{12}$ $h^{-1}$ M$_\odot$ (upper panel),
$10^{10.5}$ $h^{-1}$ M$_\odot$  (middle panel)
and $10^{9.5}$ $h^{-1}$ M$_\odot$ (lower panel).
We have plotted the curves for experiments
performed without including energy feedback (solid lines)
and including feedback with an input parameter of $\epsilon_c=0.5$ (dashed lines).}
\label{sfr}
\end{figure}

We analise here results for isolated disc galaxy simulations, in the virial mass
range $10^{9.5}-10^{12}$ $h^{-1}$ M$_\odot$. In Fig.~\ref{sfr} we show the
evolution of the star formation rate for tests of 
$10^{12}$ $h^{-1}$ M$_\odot$ (upper panel),
$10^{10.5}$ $h^{-1}$ M$_\odot$  (middle panel)
and $10^{9.5}$ $h^{-1}$ M$_\odot$ systems (lower panel).
We have plotted the curves for experiments
performed without including energy feedback (solid lines)
and including feedback with an input parameter of
$\epsilon_c=0.5$ (dashed lines).
These trends indicate that the inclusion of the model
of SN energy feedback succeeds in regulating the star
formation activity in galaxies. 
This is the result of the effects of the injection
of energy into the interstellar gas: when SNe explosions take
place, the gas is efficiently heated up and accelerated
outwards. As a result, the cold gas density from
which stars are formed decreases and  the star formation rate is reduced.
Note also that
lower mass systems are more strongly affected, as expected
since their potential wells are less efficient in retaining
baryons (Larson 1974).

\begin{figure}  
{\hspace*{2.5cm}   \psfig{figure=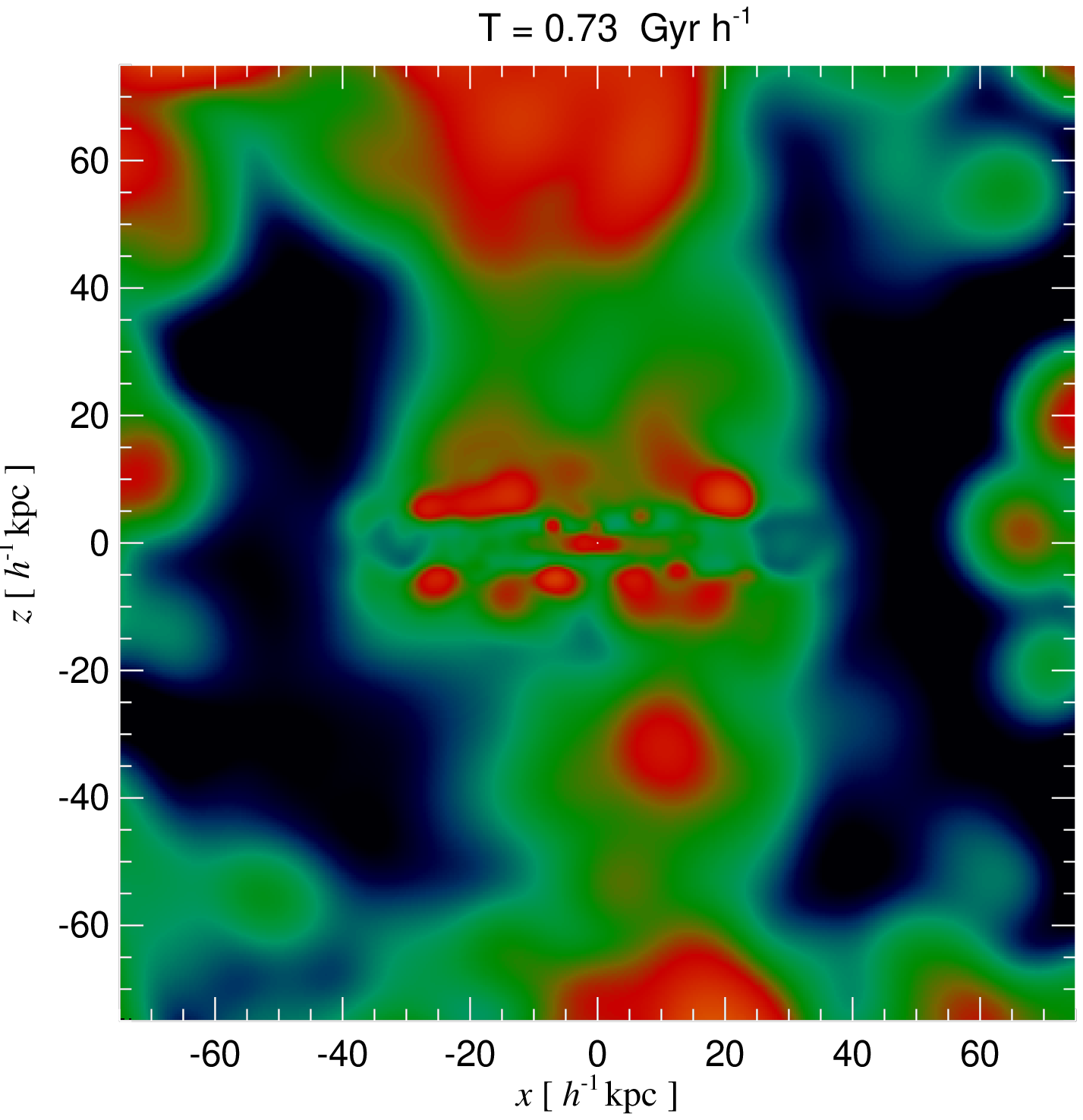,width=9.cm}}

{\hspace*{2.5cm}   \psfig{figure=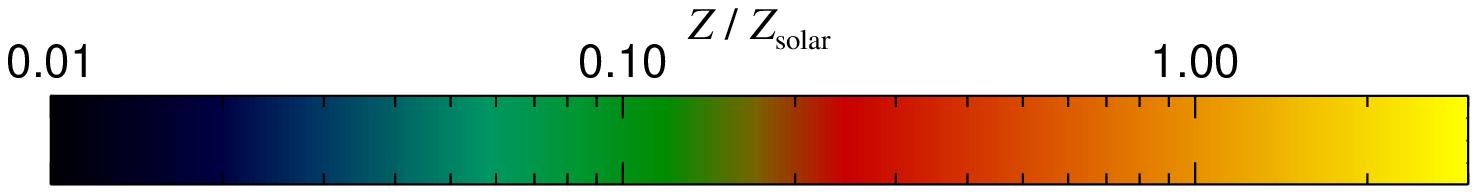,width=9.cm}}
\caption{Edge-on gaseous metallicity distribution for our  $10^{12}$ $h^{-1}$ M$_\odot$ virial
mass system run including the feedback model 
with an input parameter of $\epsilon_c=0.5$ (dashed lines), 
 after $0.75$ Gyr $h^{-1}$ of evolution. The metallicity scale is also shown.}
\label{metallicity_map}

\end{figure}

As it was mentioned before, SN feedback is responsible of heating
up the gas in star forming regions and, consequently, of
generating important outflows of material. Since these outflows
originate in the center of the systems where the gas
is highly enriched with metals, they can transport an important
fraction of the chemical elements into the haloes and even beyond.
This could in principle explain the observed presence of metals
in the intergalactic medium.
As an example, we show in Fig.~\ref{metallicity_map} the edge-on projected
gas metallicity distribution for our test of $10^{12}$ $h^{-1}$ M$_\odot$ virial
mass system just after the starburst.
From this figure we can see that our model is able to enrich the regions
outside the centre of galaxies. Also note that the generated  outflows are mostly perpendicular
to the disc plane.
 
\section{Conclusions}

We have presented results of the effects of SN feedback on the
evolution of isolated disc galaxies of different
mass, by using an extended version of the code {\small GADGET-2}
which includes a treatment of chemical and energy feedback 
by Supernova. We have shown that the model succeeds in
generating a self-regulated star formation process, as a consequence
of the triggering of strong outflows of material.

We found that the impact of SN feedback on galaxy formation
depends on virial mass. Smaller systems are more
strongly affected since they have shallower potential
wells and hence are less efficient in retaining
baryons. In particular, we found that the
star formation rates of small systems such as dwarf
galaxies can show a series of starbursts, in consistency
with recent observational findings (Kauffmann et al. 2003; Tolstoy et al. 2003).

In our model, galactic outflows transport an important fraction
of the metal content from the centre to the outskirts of galaxies.
This is of crucial importance since it is observed that the
intergalactic medium is contaminated with heavy metals. SN driven
outflows such as those generated by our model constitute a natural explanation
for this observational result.

\acknowledgments 
This work was partially supported  by the European Union's ALFA-II
program, through LENAC, the Latin American European Network for
Astrophysics and Cosmology. Simulations were run on Ingeld and HOPE PC-clusters at Institute
for Astronomy and Space Physics. We acknowledge support from
Consejo Nacional de Investigaciones Cient\'{\i}ficas y T\'ecnicas,
Agencia de Promoci\'on de Ciencia y Tecnolog\'{\i}a,  Fundaci\'on Antorchas
and Secretar\'{\i}a de Ciencia y 
T\'ecnica de la Universidad Nacional de C\'ordoba.
The authors thank the Aspen Center for Physics where part of the discussions
of this work took place.

\end{document}